\documentclass[amsmath,amssymb,amsfonts,aps,prl,twocolumn,superscriptaddress,nofootinbib]{revtex4-2}
\usepackage{float}
\usepackage{graphicx}
\usepackage{dcolumn}
\usepackage{bm}
\usepackage[utf8]{inputenc}
\usepackage[english]{babel}
\usepackage{amsfonts}
\usepackage{hyperref}
\usepackage{physics}
\usepackage{epstopdf}
\usepackage{mathtools}
\usepackage[capitalise]{cleveref}
\usepackage[space]{grffile}
\usepackage{color}
\usepackage[normalem]{ulem}
\usepackage{subfigure}

\begin{document}

\title{Entanglement Spectrum Dynamics as a 
Probe for Non-Hermitian Bulk-Boundary Correspondence in Systems with Periodic Boundaries}
\author{Pablo Bayona-Pena}
\affiliation{Center for Gravitational Physics and Quantum Information, Yukawa Institute for Theoretical Physics, Kyoto University, Kyoto 606-8502, Japan}
\author{Ryo Hanai}
\affiliation{Center for Gravitational Physics and Quantum Information, Yukawa Institute for Theoretical Physics, Kyoto University, Kyoto 606-8502, Japan}
\author{Takashi Mori}
\affiliation{ Department of Physics, Keio University, 3-14-1 Hiyoshi, Yokohama, 223-8522, Japan}
\author{Hisao Hayakawa}
\affiliation{Center for Gravitational Physics and Quantum Information, Yukawa Institute for Theoretical Physics, Kyoto University, Kyoto 606-8502, Japan}
\date{\today}

\begin{abstract}
It has recently been established that open quantum systems may exhibit a strong spectral sensitivity to boundary conditions, 
known as the non-Hermitian/Liouvillian skin effect (NHSE/LSE), making the topological properties of the system boundary-condition sensitive. 
In this Letter, we ask the query: Can topological phase transitions of open quantum systems, captured by \textit{open} boundary conditioned invariants, be observed in the dynamics of a system under \textit{periodic} boundary conditions,  even in the presence of NHSE/LSE? 
We affirmatively respond to this question,
by considering the quench dynamics of entanglement spectrum in a periodic open quantum fermionic system. 
We demonstrate that the entanglement spectrum exhibits zero-crossings only when this \textit{periodic} system is quenched from a topologically trivial to non-trivial phase, defined from the spectrum under \textit{open} boundary conditions,
even in systems featuring LSE. 
Our results reveal that non-Hermitian topological phases leave a distinctive imprint on the unconditional dynamics 
within a subsystem of fermionic systems.  
\end{abstract}

\maketitle

{\it Introduction}.---
The recent expansion of phases of matter 
into the non-Hermitian realm has significantly broadened our understanding of free fermionic systems~\cite{PRXGong18, Kawabata19, Shen18}. 
The presence of non-Hermiticity, which generally arises from the exchange of energy and/or particles of a system with the environment, leads to various exotic
phenomena. A paradigmatic example is the emergence of
extreme spectral sensitivity to boundary conditions, often referred to as the \textit{non-Hermitian skin effect} (NHSE)~\cite{Alvarez18, Yao18, Yokomizo19, YaoSong19, Haga21, Mori20, Hamanaka23, Lee23, McDonald22, Sam24, Fang22, Imura19, Yang20, Okuma20, Borgnia20, Lei20, Shen23, Gideon23, Longhi20, Okuma21, Mao21}, or \textit{Liouvillian skin effect} (LSE) in the context of open quantum systems~\cite{YaoSong19, Haga21, Lee23, McDonald22, Sam24, Fang22}, see also~\cite{Mori20, Hamanaka23} for related results in interacting many-body systems. 
In the presence of NHSE/LSE, the system exhibits exponential localization of eigenstates near its boundaries. This poses a challenge to the conventional understanding of bulk-boundary correspondence: the presence of topological edge states in open boundary conditions (OBC) cannot be inferred from the topological invariant of the Bloch Hamiltonian~\cite{Yao18,Yokomizo19}. In this situation, one must extend the conventional Bloch band theory to incorporate complex-valued wave vectors in order to identify topological invariants in OBC~\cite{Yao18,Yokomizo19}. This is known as the non-Hermitian bulk-boundary correspondence.

Let us remark, however, a somewhat obvious point that
for systems subject to periodic boundary conditions (PBC), 
the topological transition point 
is determined by the Bloch Hamiltonian
even in the presence of NHSE/LSE. It is tempting to expect that no sign of topological invariants defined from the OBC spectrum can be extracted from the PBC system.\par
In this Letter, in sharp contrast to this na\"ive expectation, we show that there are cases where OBC topological invariants can be extracted from the dynamics of the PBC system. 
This is achieved by examining the quench dynamics of the entanglement spectrum (ES).
Initially recognized in the context of 
a fractional quantum Hall insulator~\cite{Li08} and later formalized for free fermionic~\cite{Fidkowski10} and one-dimensional interacting systems~\cite{Pollman10},
the ES has since played a pivotal role in understanding topological phases of both non-interacting and interacting systems. In highly non-equilibrium states, the ES serves as a dynamical probe for wavefunction topology~\cite{Gong18, McGinley18, Wang17}.
Recent investigations have partially extended its applicability to non-Hermitian and open quantum systems. In non-Hermitian systems, which effectively describe the dynamics of open systems subject to post-selection, it was reported that for point-gapped systems, which exhibit NHSE, ES fails to capture topological properties of the ground state~\cite{Loic19}.
In open quantum systems, the real line-gap topology
was shown to be detectable
through the quench dynamics of ES~\cite{Sayyad21,Starchl22, Starchl24}. However, these works on open quantum systems, considered systems that do not exhibit LSE~\cite{Sayyad21, Starchl22, Starchl24}, leaving the effect of boundary conditions sensitivity on the dynamics of ES unclear.

We examine in this Letter an open quantum fermionic lattice exhibiting LSE and show that the dynamics of the ES can serve as a probe for topological phases defined from the OBC spectrum, even when the physical system is in PBC. In particular, we numerically analyze an open quantum model whose dynamics is governed by a dynamical matrix identical to the non-Hermitian Su-Schrieffer-Heeger (SSH) model, a paradigmatic model exhibiting NHSE/LSE~\cite{Yao18,YaoSong19,Yokomizo19,McDonald22,Fang22, Fang22}
and demonstrate that in PBC systems, topological phases defined from the OBC spectrum of Lindbladians are discernible through mode crossings of the ES during quench dynamics. As the ES in open free fermionic systems can be computed from a two-point correlation function of a  subsystem~\cite{Ingo03},
our proposal to detect the non-Hermitian bulk-boundary correspondence does not require post-selection.\par
We remark that conventional observables such as the correlation functions and steady-state density profile are \textit{not} generically affected by NHSE/LSE, 
\cite{Longhi20, Okuma21, Mao21, Fang22, McDonald22, Lee23, Fang22, Sam24}.
Intuitively, this is due to the locality of the system: the system's response to an external perturbation applied in the bulk is unaffected by the boundary condition unless the effect reaches the boundary before it dissipates.
Our finding here seems to contradict these previous results, but it does not:
the ES
is a highly non-local quantity
(for macroscopically large subsystem sizes that we consider in this Letter), 
making it possible to catch the (non-local) spectral sensitivity to boundary conditions stemming from LSE.

\begin{figure}[t]
    \center
    \includegraphics[width= 0.23\textwidth]{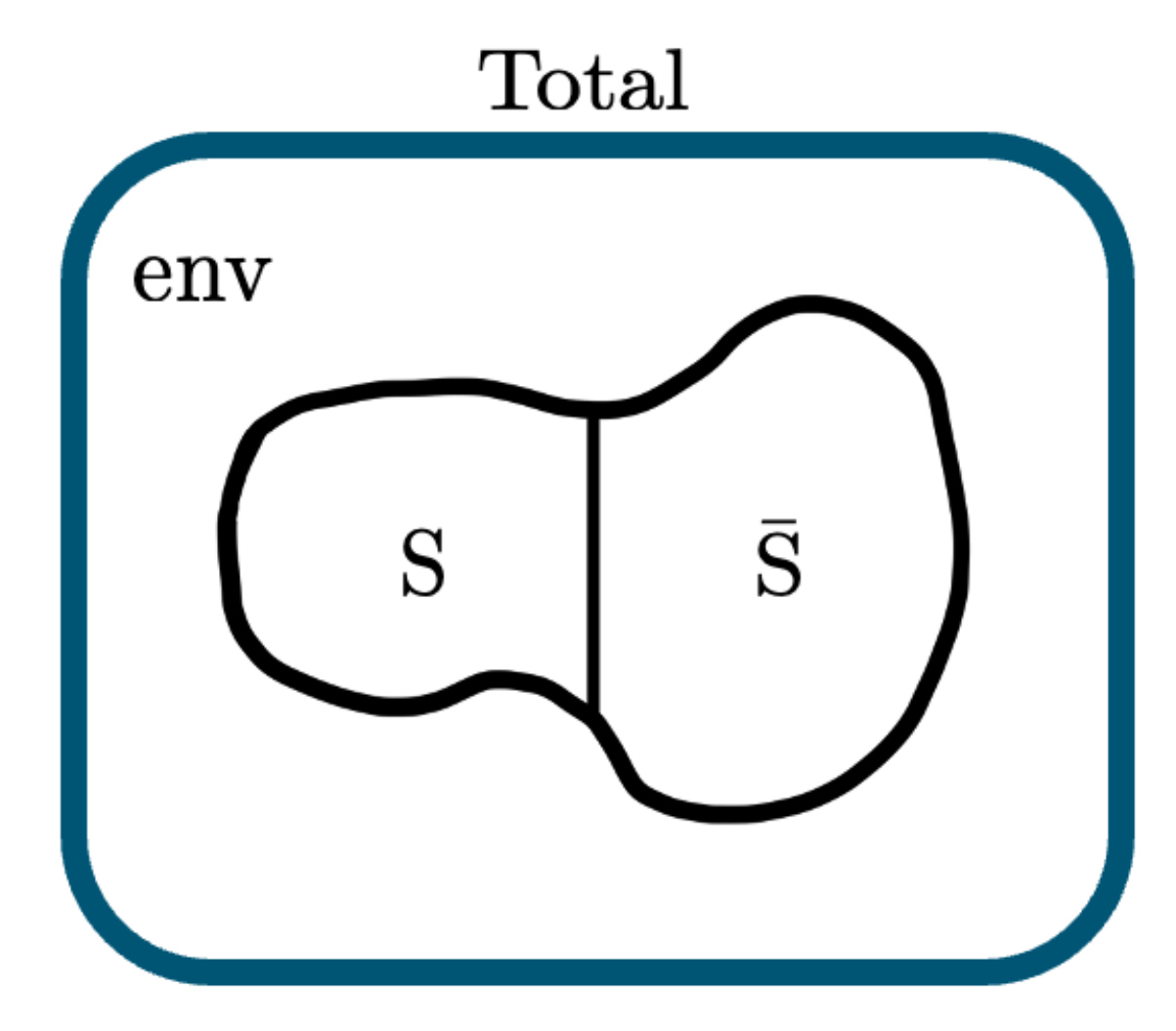}
    \label{f0}
    \caption{Illustration of tri-partition of an open quantum system. The bipartite relevant system ($\textrm{S}+\bar{\textrm{S}}$) is allowed to interact with the environment (env). We focus on the coarse-grained dynamics of the relevant system after tracing out the environmental degrees of freedom $\hat{\rho}= {\rm Tr}_{\rm env}[\hat{\rho}_{\rm tot}]$.
    }
    \label{fig_tripartition}
\end{figure}
\begin{figure*}[t]
    \center    \includegraphics[width=0.78\textheight]{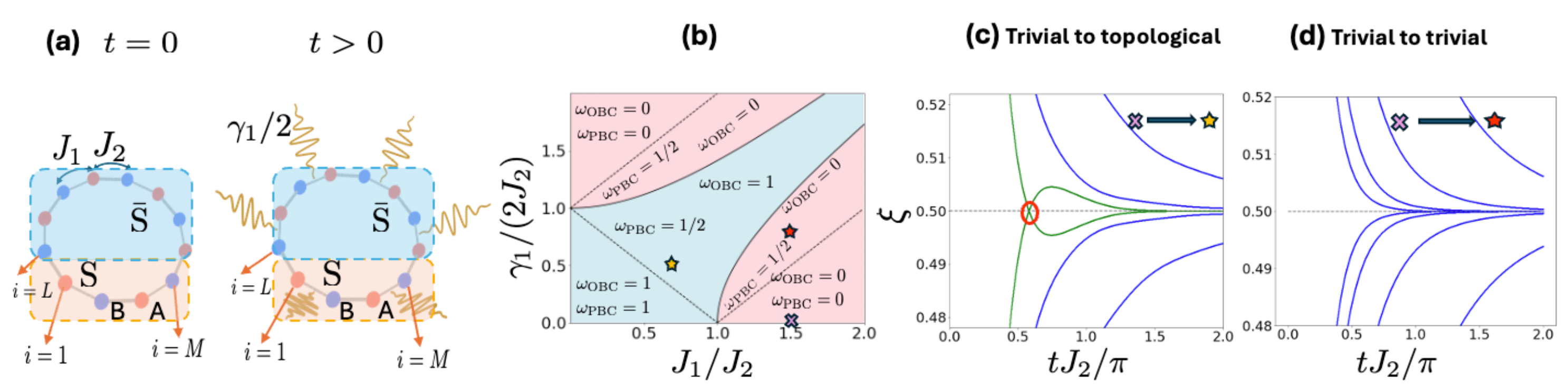}
    \caption{
        (a) The system is initially prepared in the trivial ground state of the SSH model ~\cite{SSH79} under PBC
        with no dissipation.
        Then, at $t >0$, the relevant system $(\mathrm{S}+\bar{\mathrm{S}})$ is driven out of its initial state by an abrupt change of the hopping parameters $J_1, J_2$ and a correlated dissipation with rate $\gamma_1$ is simultaneously turned on. The relevant system evolves according to the GKLS equation Eq.~\eqref{quadratic_Lind}
        under PBC.
        (b) Phase diagram of the Lindbladian SSH model Eq.~\eqref{targetH}. Topological phase transitions are separated by the solid and dotted lines for OBC and PBC, respectively. 
        (c), (d) ES dynamics for a subsystem S. (c) 
        Trivial to topological quenches for both OBC and PBC spectrum ($\omega_{\rm OBC}=1, \omega_{\rm PBC}=1/2$)
        $(J_1/J_2,\gamma_1/(2J_2))=(0.7,0.5)$. 
        The green lines represent the four eigenvalues, $\xi_l(t)$, of the covariance matrix $\mathsf{C}_{\rm SS}(t)$ that are closest to the zero energy entanglement value $\xi=0.5$ during the time evolution. The other eigenvalues are shown in blue. Four ES eigenvalues 
        cross $\xi$ = 0.5 as they decay. 
        (d) 
        Trivial to trivial
        quenches for OBC spectrum ($\omega_{\rm OBC}=0$)~$(J_1/J_2,\gamma_1/(2J_2))=(1.5,0.8)$ but topological for PBC spectrum ($\omega_{\rm PBC}=1/2$).  All the eigenvalues, $\xi_l(t)$, of $\mathsf{C}_{\rm SS}(t)$ are shown in blue. All eigenvalues decay to $\xi=0.5$ without crossing it.  For both cases, we set the initial state to be the (trivial) ground state at $(J_1/J_2, \gamma_1/(2J_2))=(1.5,0)$,
        the system size $L=20$, and the size of subsystem S is $M=10$ sites.
        Note crucially that, while the ES dynamics reflect the OBC spectrum, our physical system is under PBC.
        }
    \label{fig_LSSH}
\end{figure*}

\textit{The model.}---
We investigate the dynamics of an open quantum system, illustrated in Fig.~\ref{fig_tripartition}, where a free fermionic system ($\mathrm{S}+\bar{\mathrm{S}}$) is linearly coupled to a Markovian environment (${\rm env}$). After integrating out the environmental degrees of freedom,
the reduced density matrix $\hat{\rho}= {\rm Tr}_{\rm env}\hat{\rho}_{\rm tot}$ is governed by
a quadratic Gorini-Kossakowski-Sudarshan-Lindblad (GKSL) equation~\cite{Lindblad76,GKS76}, given by,
\begin{align}
    \partial_t \hat{\rho}(t)  &=-i[ \hat{H}, \hat{\rho}(t)] + \sum_{\alpha }\mathcal{D}[\hat{L}_\alpha]\hat{\rho}(t)
    + \sum_{\alpha }\mathcal{D}[\hat{G}_\alpha]\hat{\rho}(t) \nonumber \\
    & =
     \hat{\mathcal{L}}\hat{\rho}(t),
    \label{quadratic_Lind}
\end{align}
where the superoperator $\hat{\mathcal{L}}$ is  the Liouvillian and $\hat{H}= \sum_{i,j=1}^L H_{i,j} \hat{c}_i^\dagger \hat{c}_j$ is the Hamiltonian of the relevant system $(\textrm{S}+\bar{\textrm{S}})$. Here, $\hat{c}_i^\dagger$ and $\hat{c}_i$
are creation and annihilation operators of fermions at site $i$, respectively. 
The dissipation arising from
the coupling of the system to the Markovian reservoir is described by $\mathcal{D}[\hat{M}]\hat{\rho}:= \hat{M}\hat{\rho} \hat{M}^\dagger -\frac{1}{2} \{ \hat{M}^\dagger \hat{M}, \hat{\rho}(t) \}$. 
Each coupling to the Markovian reservoir, labeled by $\alpha$ (see Fig.~\ref{fig_LSSH}(a)), induces incoherent addition or removal of particles and is described by linear jump operators,  
$\hat{L}_{\alpha}= \sum_i \ell_{ i}^\alpha \hat{c}_i$ and $\hat{G}_{\alpha}= \sum_i g^{* \alpha}_{i}\hat{c}^\dagger_i$,  respectively,  
where $\ell_{i}^\alpha$ and $g^\alpha_i$ are complex numbers. 

According to the tenfold way classification of quadratic fermionic Lindbladians~\cite{Lieu20}, the entire topology of a quadratic Lindbladian can be characterized by the topological invariants of a non-Hermitian operator $\hat{H}_{\rm eff}$~\cite{McDonald22}:
\begin{equation}
    \hat{ H}_{\rm eff}:= {\hat H} -\frac{i}{2}({\hat L}+{\hat G}) ,
    \label{Heff}
\end{equation}
where $ \hat{L}= \sum_\alpha \hat{L}_\alpha^\dagger \hat{L}_\alpha,$ and $\hat{G}= \sum_\alpha \hat{G}_\alpha \hat{G}_\alpha^\dagger$. 
We note that $\hat H_{\rm eff}$ is different from the conditional Hamiltonian $\hat H_{\rm cond}=\hat H - (i/2)(\hat L - \hat G)$ that governs the dynamics of null-jump processes~\cite{YaoSong19, McDonald22, Lee23, Sam24, Pablo24}.
\par
\textit{ES crossings as a 
non-Hermitian bulk-boundary correspondence indicator}.---
To dynamically detect the non-Hermitian bulk-boundary correspondence, we propose examining the ES
dynamics after a dissipative quench. 
As depicted in Fig.~\ref{fig_tripartition}, we partition the relevant system $(\mathrm{S}+\bar{\mathrm{S}})$ further into the subsystem ${\mathrm{S}}=\{1, 2, ..., M\}$ and its complement $\bar{\mathrm{S}}=\{M+1, M+2, ..., L\}$. 
The reduced density matrix of the subsystem can be written as $\hat{\rho}_\mathrm{S}={\rm Tr}_{\bar{\mathrm{S}}}\hat{\rho}=(1/Z_{\mathrm{S}})e^{-\hat H_{\rm S} }$ with $Z_{\mathrm{S}}={\rm Tr}_{\mathrm{S}}[e^{-\hat H_{\rm S} }]$, where ${\rm Tr}_{\mathrm{S}(\bar{\mathrm{S}})}$ denotes the trace over subsystem ${\mathrm S}(\bar{\mathrm{S}})$.
We refer to the spectrum of $\hat{H}_{\rm S}$ as the ES $\epsilon_l$, the central quantity of interest. Using the property that $\hat\rho_{\rm S}$ is Gaussian, the entanglement Hamiltonian can be decomposed as $\hat H_{\rm S}= \sum_l \epsilon_l \hat{f}_l^\dagger \hat f_l$, where  $\hat f_l$ is a fermionic annihilation operator of the  $l$-th eigenmode.
In the absence of the system-environment coupling, this recovers the conventional definition of the ES for closed fermionic systems, which is known to successfully detect bulk-boundary correspondence~\cite{Li08,Fidkowski10,Pollman10}.\par
References~\cite{Gong18, Wang17, McGinley18}
considered the situation where
they first prepared a state corresponding to the ground state of a Hamiltonian in the trivial phase followed by a sudden change in the parameters. 
They showed that when the post-quench Hamiltonian has a topologically non-trivial (trivial) ground state, (no) zero-crossings of the ES occur.\par
We investigate below a similar setup where we also quench the system from a topologically trivial to a non-trivial or trivial state in an open quantum system governed by Eq.~\eqref{quadratic_Lind}~\cite{Sayyad21, Starchl22, Starchl24}. Importantly, the topology of our system is characterized by the non-Hermitian effective Hamiltonian $\hat{H}_{\rm eff}$, given by Eq.~\eqref{Heff}, that exhibits NHSE.
We find that an analogous phenomenon to the closed-system counterpart occurs, which surprisingly 
reflects
the topology of an 
\textit{OBC} system even when the system itself is in 
\textit{PBC}: 
physically traversing the system is unnecessary to detect topological phases of
$\hat{H}_{\rm eff}$ defined under open boundary conditions.

Practically, in our quadratic system,
the ES can be 
computed from the covariance matrix $\mathsf{C}(t)$,
whose $(j,i)$ component is given by the spatial correlation function
$[{\sf C}]_{j,i}(t):= {\rm Tr}_{\mathrm{S}+\bar{\mathrm{S}}}[\hat{\rho}(t) \hat{c}^\dagger_i \hat{c}_j]$ \cite{Ingo03}. 
By taking advantage of the property that the density matrix is always Gaussian  (as long as we take the initial state to be Gaussian) and Wick's theorem applies as a result, 
the ES can be derived from the spectrum of the covariance matrix ${\sf C}_{\sf SS}$ in the subsystem $\textrm{S}$ defined as the correlation matrix $\sf C$ whose indices are restricted to subsystem S,
\begin{equation}
    [{\sf C}_{\rm SS}]_{j,i}(t) :=  
   {\rm Tr}_{\mathrm{S}+\bar{\mathrm{S}}}[\hat{\rho}(t) \hat{c}^\dagger_i \hat{c}_j];\; (i,j) \in {\sf S}, 
\end{equation}
which can be expressed using the reduced density matrix of the subsystem S as $ [{\sf C}_{\rm SS}]_{j,i}(t)= {\rm Tr}_{\rm S}[\hat{\rho}_{\rm S}(t) \hat{c}^\dagger_i \hat{c}_j]$. 
Letting
$\xi_l(t)$ be the eigenspectrum of $\mathsf{C}_{\rm SS}(t)$, 
one finds the relation between $\xi_l(t)$ and the ES $\epsilon_l(t)$
~\cite{Ingo03} (See Supplemental Materials 
(SM)~\cite{Pablo24}):
\begin{equation}
    \xi_l(t)= \frac{1}{e^{\epsilon_l(t)}+1}.
    \label{SingleParticleES}
\end{equation}
An entanglement zero mode with $\epsilon_l = 0$ corresponds to $\xi_l = 0.5$. 
To investigate the dynamics of $\xi_l$, we derive from Eq.~(\ref{quadratic_Lind}) the following equation for the covariance matrix dynamics~\cite{McDonald22, Sam24}:
\begin{equation}
    i \partial_t  \mathsf{C}(t)= \mathsf{H}_{\rm eff} \mathsf{C}(t) -\mathsf{C}(t) \mathsf{H}_{\rm eff}^\dagger+ i\mathsf{G},
    \label{C_dynamics}
\end{equation}
where $[\mathsf{G}]_{i,j}=\sum_{\alpha} g^{*\alpha}_i g^\alpha_j $ and $[\mathsf{H}_{\rm eff}]_{i,j}= H_{i, j} -i/2\sum_\alpha(\ell^{*\alpha}_i \ell^\alpha_j +g^{*\alpha}_i g^\alpha_j )$. $g^\alpha_i$ and $\ell_i^\alpha$ are complex numbers. 
\par
\textit{Lindbladian SSH model}.---
To be specific, we consider the following system
exhibiting LSE \cite{Yao18,Fang22} (Fig.~\ref{fig_LSSH}(a)). For the system Hamiltonian, we consider the tight-binding Hamiltonian
\begin{equation}
     \hat{H} = \sum\limits_{j} \left( J_1 \hat{c}_{A,j}^\dagger 
     \hat{c}_{B,j} + J_2 \hat{c}^\dagger_{B,j} \hat{c}_{A,j+1}+ 
     J_3 \hat{c}_{A,j}^\dagger
     \hat{c}_{B,j+1} \right) + h.c. \label{SSH} \end{equation}
For $J_3=0$, the Hamiltonian reduces to the SSH model~\cite{SSH79}. The dissipation of our model is given by the following jump operators for $j=1,2,\cdots, N$ (see Fig.~\ref{fig_LSSH} (a)): 
   $ \hat{L}_{j,1}  = \sqrt{\frac{\gamma_1}{2} } \left( \hat{c}_{A,j} +i \hat{c}_{B,j} \right)$,
    $\hat{G}_{j,1}  = \sqrt{ \frac{\gamma_1}{2}} (\hat{c}_{A,j}^\dagger -i\hat{c}_{B,j}^\dagger)$,
   $\hat{L}_{j,2}  = \sqrt{ \frac{\gamma_2}{2} } \left(\hat{c}_{B,j}+ i\hat{c}_{A,j+1}   \right)$,
   $\hat{G}_{j,2} = \sqrt{ \frac{\gamma_2}{2}} 
   (\hat{c}_{B,j}^\dagger-i\hat{c}_{A,j+1}^\dagger )$. Here, we have chosen jump operators such that $\hat L_{j,s=1,2}^\dagger=\hat G_{j,s=1,2}$, ensuring
the system reaches an infinite-temperature state at the long-time limit. 
The ES asymptotically converges to zero $\epsilon_n=0$ $(\xi_n=0.5)$, avoiding the ES from crossing zero for trivial reasons with no topological origin~\cite{Sayyad21, NOTE}.\par
This choice of Hamiltonian and jump operators 
yield the effective Hamiltonian (Eq.~\eqref{Heff}), 
\begin{align}
   & \hat{H}_{\rm eff}  =  \sum \limits_{j} \Big[ \Big(J_1 \hat{c}_{A,j}^\dagger \hat{c}_{B,j} + J_2 \hat{c}^\dagger_{B,j}
   \hat{c}_{A,j+1} \nonumber\\
   &+ J_3 \hat{c}_{A,j}^\dagger \hat{c}_{B,j+1} \Big) + h.c. \Big]  + \sum_j \Big( \frac{\gamma_1}{2}( \hat{c}_{A,j}^\dagger \hat{c}_{B,j} -\hat{c}^\dagger_{B,j} \hat{c}_{A,j} ) \nonumber \\
 &+\frac{\gamma_2}{2} ( \hat{c}^\dagger_{B,j} \hat{c}_{A,j+1} -\hat{c}_{A,j+1}^\dagger \hat{c}_{B,j})\Big) -i \frac{\gamma_1+\gamma_2}{2} \hat{n} ,
    \label{targetH}
\end{align}
with $\hat{n}:= \sum_j [\hat{c}^\dagger_{A,j} \hat{c}_{A,j}+ \hat{c}^\dagger_{B,j} \hat{c}_{B,j}]$,
which is the non-Hermitian SSH model well studied in the literature (except for the constant shift in the imaginary axis that does not affect its topological features) which is known to exhibit NHSE~\cite{Yao18,Yokomizo19}. For PBC, 
the Hamiltonian $\hat{H}_{\rm eff} $ can 
expressed as $\hat{H}_{\rm eff}= \sum_k \boldsymbol{c}^\dagger(k) \mathsf{H}_{\rm eff}(k) \boldsymbol{c}(k)$, where $\boldsymbol{c}(k) := (\hat{c}_A(k), \hat{c}_B(k))^T$ with
$\hat{c}_{A(B)}(k)= 1/\sqrt{L}\sum_j e^{-ikj}\hat{c}_{A(B),j} $ and $\mathsf{H}_{\rm eff}(k)= h_1(k)\sigma_x +h_2(k)\sigma_y$, where $h_1(k)= J_1 +(J_2+J_3)\cos(k) -i\gamma_2/2 \sin(k)$, $h_2(k)= (J_2-J_3)\sin(k)+ i\gamma_2/2 \cos(k)-i\gamma_1/2 $. Here,
$\sigma_i$ with $i\in \{x,y,z \}$ are the Pauli matrices and we have ignored the constant shift term.
Since the effective Hamiltonian has 
a sublattice symmetry $\sigma_z \mathsf{H}_{\rm eff}(k)\sigma_z^{-1}=- \mathsf{H}_{\rm eff}(k)$,
we can define the  ${\mathbb{Z}}$ topological number of $\mathsf{H}_{\rm eff}(k)$, given by the winding number $\omega_{\rm PBC}=\oint \frac{dk}{4\pi i} {\rm Tr}[\sigma_z \mathsf{H}_{\rm eff}^{-1}(k) \frac{d}{dk}\mathsf{H}_{\rm eff}(k)]$~\cite{Kawabata19}. 
The winding number can be similarly defined for OBC, $\omega_{\rm OBC}$, by 
generalizing the momentum integral in the PBC winding number, $\omega_{\rm PBC}$, to those with complex values along the generalized Brillouin zone. (See e.g., Refs.~\cite{Yokomizo19, Yang20} for details.)\par
First, we focus on the case where $\gamma_2=0,\;J_3=0$. Here,
in PBC, the topological transition point is given by $J_1= J_2 \pm \gamma_1/2 \;$ or $J_1= -J_2 \pm \gamma_1/2 \;$~\cite{Yao18}. 
For OBC, 
the spectrum can be easily obtained by means of a similarity transformation $\mathsf{S}$, such that $\mathsf{H}_{\rm eff}$ and  $\bar{\mathsf{H}}_{\rm eff}:= \mathsf{S}^{-1}\mathsf{H}_{\rm eff}\mathsf{S}$ share the same spectrum. 
By taking $\mathsf{S}={\rm diag}(1,r,r,r^2,r^2,\cdots, r^{L-1},r^{L-1},r^{L})$
with $r=\sqrt{| (J_1 - \gamma_1/2)/(J_1 + \gamma_1/2)|}$~\cite{Yao18, Yokomizo19}, 
$\mathsf{\bar{H}}_{\rm eff}$ becomes the Hermitian SSH model with hopping parameters $\bar{J}_1 =\sqrt{(J_1 - \gamma_1/2)(J_1 + \gamma_1/2)}$ and $\bar{J}_2= J_2$
provided that $|J_1|>|\gamma_1/2|$.
The phase boundary between topological and trivial phases 
can then be obtained as (assuming $J_1>0$ without loss of generality):
\begin{align}
    J_1 &= \sqrt{J_2^2 +\left(\frac{\gamma_1}{2}\right)^2} 
    \;\;\;\;{\rm or} \;\;\; \sqrt{-J_2^2 +\left(\frac{\gamma_1}{2}\right)^2},
    \label{ciritcal_z}
\end{align} 
as shown in Fig.~\ref{fig_LSSH}(b). 
The discrepancy between OBC and PBC phase diagrams is due to the NHSE as shown in Fig.~\ref{fig_LSSH}(b)~\cite{Yao18, Yokomizo19, Kawabata19}.

We numerically demonstrate below that the topological transition point in OBC given by Eq.~\eqref{ciritcal_z} can be detected through the ES dynamics of a PBC system after a quench.
\begin{figure}[t]
    \center
    \includegraphics[width= 0.45\textwidth]{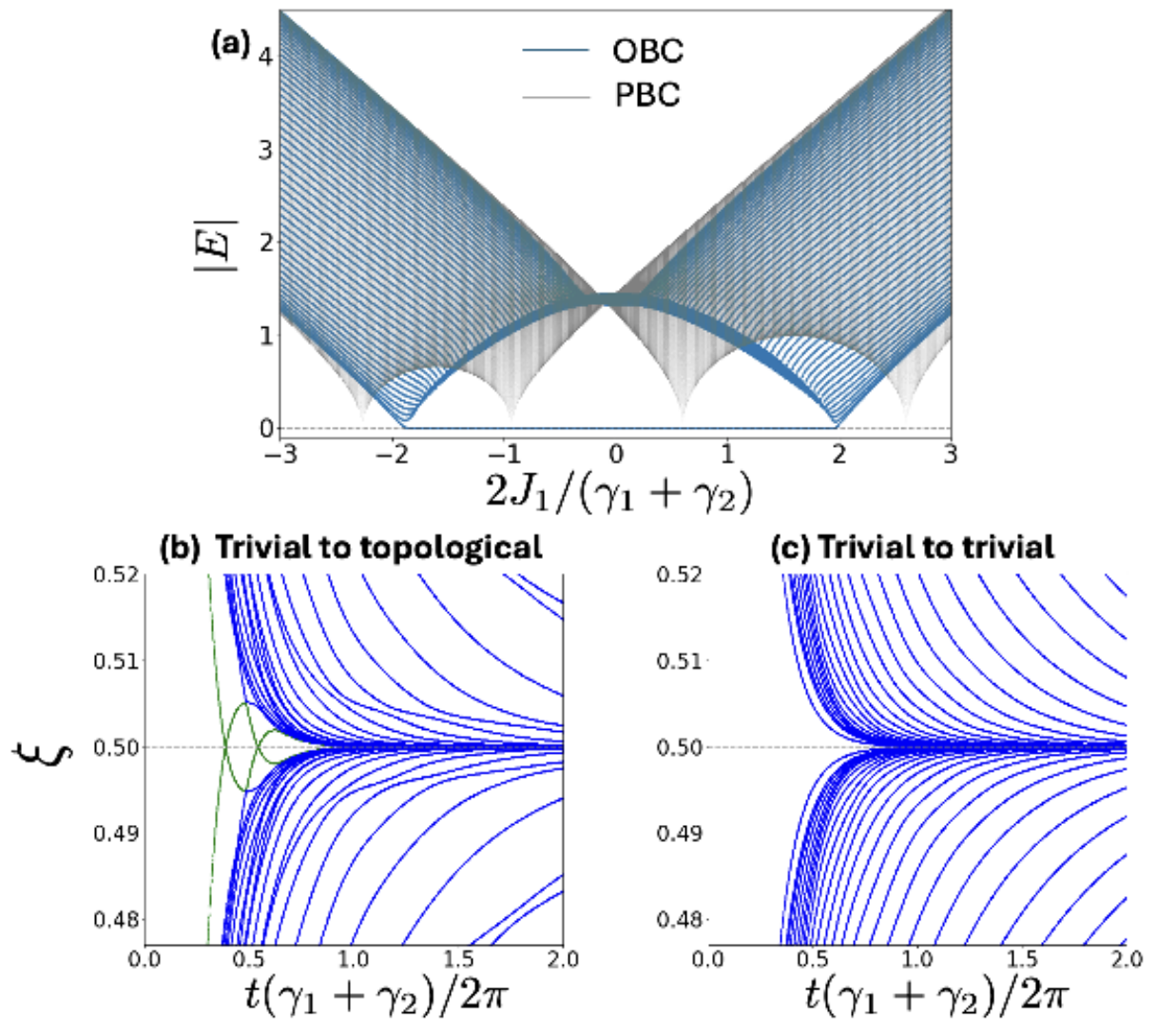}
    \caption{
    (a) 
    The blue (grey) lines are the energy bands of the effective Hamiltonian $\mathsf{H}_{\rm eff}$
    in a finite open chain for 
    $2J_2/(\gamma_1+\gamma_2)=1.4,\;  2J_3/(\gamma_1+\gamma_2)= 1/5,\; 2\gamma_1/(\gamma_1+\gamma_2)= 5/3,\; 2\gamma_2/(\gamma_1+\gamma_2)=1/3$  
    in OBC (PBC).
    (b),(c) Quench dynamics of ES under PBC. The system is initially prepared in the trivial ground state of the SSH model $(J_2/J_1, J_3/J_1, \gamma_1/J_1,\gamma_2/J_1)= (0,0,0,0)$, 
    with system size $L$ = 40, and a subsystem S
    of $M=20$ sites. At $t=0$ the relevant system $(\mathrm{S}+\bar{\mathrm{S})}$ is driven out of equilibrium by an abrupt change in parameters and allowing the system to couple to Markovian reservoirs at a rate $\gamma_i, (i=1,2)$. (b) Trivial to topological quench for OBC and PBC spectrum $(2J_1/(\gamma_1+\gamma_2),2J_2/(\gamma_1+\gamma_2),2J_3/(\gamma_1+\gamma_2))=(-1,1.4,1/5)$. The green lines represent the four eigenvalues, $\xi_l(t)$, of the covariance matrix $\mathsf{C}_{\rm SS}(t)$ that are closest to $\xi=0.5$ during the time evolution.  The other eigenvalues are shown in blue. Two pairs of two ES eigenvalues, $\xi_l(t)$, cross $\xi=0.5$ at different times as they decay. (c) Trivial to trivial (topological) quench for OBC (PBC) spectrum $(2J_1/ (\gamma_1+\gamma_2),2J_2/(\gamma_1+\gamma_2),2J_3/(\gamma_1+\gamma_2))=(2.1,1.4,1/5)$. All the eigenvalues $\xi_l(t)$ shown in blue decay without zero-crossings. 
    }
    \label{fig_NextNearest}
\end{figure}
In our simulation, we set the pre-quench state to be the ground state in the trivial phase of Eq.~\eqref{targetH}, in the Hermitian limit $\gamma_1=\gamma_2=0$ with $J_3=0$.

Figure \ref{fig_LSSH}(c) and (d) show the quench dynamics of ES for a PBC system.
Here, in panel (c), the post-quench parameters are in the topologically non-trivial phase for both OBC and PBC.
In panel (d), on the other hand, the parameters are chosen such that the post-quench system is in a trivial (non-trivial) phase for a OBC (PBC) system. 

For the trivial to topological quench in Fig.~\ref{fig_LSSH}(c), we observe four ES eigenvalues crossing exactly at the entanglement zero mode energy, $\xi=0.5$, as they decay.   
 In contrast, in Fig.~\ref{fig_LSSH}(d), we do not observe any crossings between eigenmodes, which exponentially decay towards the steady state. In the SM~\cite{Pablo24}, we have numerically confirmed that the presence and absence of  ES zero-crossings is determined by the phase boundary of the effective Hamiltonian $\mathsf{H}_{\rm eff}$ for OBC, given by Eq. ~\eqref{ciritcal_z}.
This demonstrates a remarkable feature: the ES cares only about the topology of the OBC spectrum.
We have also confirmed that the presence (absence) of ES zero-crossings is unaffected by the presence of disorder which respects the symmetries of the effective Hamiltonian ${\hat H}_{\rm eff}$ and by the strength of uniform on-site balanced gain and loss (that does not affect the topology of $\hat H_{\rm eff}$ but do give rise to a finite lifetime to the response to a local perturbation~\cite{McDonald22,Lee23,Sam24}), consistent with our expectation that the ES zero-crossings have a topological origin. 
 These findings motivate us to assert that the ES crossings have a topological origin, and their presence (or absence) is determined by the topological invariants of the pre-quench Hamiltonian ${\hat H}$ and the effective Hamiltonian ${\hat H}_{\rm eff}$.

Finally, we consider the more general case with $\gamma_2 \neq 0, J_3 \neq 0$
and demonstrate that the above features remain qualitatively intact even in cases where the effective Hamiltonian $\mathsf{H}_{\rm eff}$ cannot be transformed to a Hermitian Hamiltonian. 
Figure \ref{fig_NextNearest}(a) shows the single-particle spectrum of the effective Hamiltonian $\hat H_{\rm eff}$ for both OBC and PBC. 
Similarly to the previous case, we see that the topological transition point, marked by the closing of the energy gap, is different between OBC and PBC, again, due to the NHSE.

Figure~\ref{fig_NextNearest}(b) and (c) show the quench dynamics of ES for a PBC system, where, as before, we set the pre-quench state to be the ground state of the Hermitian SSH model in the trivial phase. In panel (b), the post-quench parameters lie in the topologically non-trivial phase for both OBC ($\omega_{\rm OBC}=1$) and PBC ($\omega_{\rm PBC}=1/2$), 
whereas for panel (c), quench parameters lie in the trivial phase for OBC ($\omega_{\rm OBC}=0$) and in the topological phase for PBC ($\omega_{\rm PBC}= -1/2$). We observe that two pairs of two ES eigenvalues, $\xi_l(t)$, crossing at $\xi=0.5$ for the trivial to topological quench shown in Fig.~\ref{fig_NextNearest}(b),
while Fig.~\ref{fig_NextNearest}(c) shows that no crossings occur for trivial to trivial quenches. 
This demonstrates that
the presence or absence of zero-crossings in the ES dynamics is determined by the OBC phase diagram of the effective Hamiltonian $\hat{H}_{\rm eff}$. In the SM~\cite{Pablo24}, we have numerically demonstrated that the findings presented in this letter hold true for non-equilibrium quenches, where the system under consideration, ${\rm S+ \bar{S}}$, evolves under OBC conditions.

We briefly note that, interestingly, the ES degeneracy at the crossing point observed in the trivial to topological phase quench, which appears in the absence of next-nearest neighbor hopping ($J_3 = 0$) [Fig.~\ref{fig_LSSH}(c)] and in the Hermitian case~\cite{Gong18}, is lifted when $J_3 \neq 0$ in our open system scenario. Additionally, we observe that for quenches on open chains, ES crossings remain degenerate. Furthermore, the possibility that ES crossings signal a topological phase transition of the entanglement Hamiltonian, $\hat{H}_{\rm S} = -\log(Z_{\rm S} \hat{\rho}_{S})$, as reported in the quench dynamics of closed systems~\cite{Gong18}, remains an open question. We currently lack an explanation for these features, leaving this as a topic for future work.

\textit{Conclusions.}---
In summary, our work has revealed that non-Hermitian bulk-boundary correspondence can be detected through the quench dynamics of
the entanglement spectrum, specifically through the presence or absence of zero-crossings.
Our results contribute to a deeper understanding of topological phases in open quantum systems.\par
\textit{Acknowledgments.}---
We thank Zongping Gong, Yuto Ashida, Kohei Kawabata, and Tianqi Chen for their helpful discussions.
P. B.-P. acknowledges financial support from the Rotary Yoneyama Foundation and the Grant-in-Aid for Scientific Research No. 21H01006.
He also thanks RIKEN for their warm hospitality during his stay there, where part of this work was initiated.
RH was supported by the Grant-in-Aid for Research Activity Start-up from JSPS in Japan (Grant No. 23K19034). 
TM acknowledges support by JSPS KAKENHI Grant Numbers JP21H05185 and by JST, PRESTO Grant No.JPMJPR2259. 
HH was supported by the Kyoto University Foundation.


\newpage
\setcounter{equation}{0}
\setcounter{figure}{0}

\renewcommand{\theequation}{S\arabic{equation}}
\renewcommand{\thefigure}{S\arabic{figure}}
\renewcommand{\bibnumfmt}[1]{[S#1]}
\renewcommand{\citenumfont}[1]{S#1}

\onecolumngrid
\clearpage

\section*{Supplemental Material}

\subsection{Conditional and Unconditional Dynamics for Quadratic Lindbladians }
In this section, we show that the dynamics of the two point correlation function is given by Eq.~(5). We also discuss the subtle difference between the non-Hermitian Hamiltonian $\hat H_{\rm cond}$ that governs the null-jump conditioned dynamics
and the effective Hamiltonian $\hat{H}_{\rm eff}$
that governs the unconditioned dynamics of the two point correlation function. The GKSL equation (1) can be rewritten as,
\begin{align}
    \partial_t \hat{\rho}(t)  &= -i[ \hat{H}, \hat{\rho}(t)] - \frac{1}{2}\sum_{\alpha } \{\hat{L}_\alpha^\dagger \hat{L}_\alpha, \hat{\rho}(t)\} -\frac{1}{2}\sum_\alpha \{ \hat{G}^\dagger_\alpha \hat{G}_\alpha , \hat{\rho}(t)\}
    +
    \sum_\alpha
    [\hat L_\alpha \hat\rho 
    \hat L_\alpha^\dagger
    +
    \hat G_\alpha \hat\rho 
    \hat G_\alpha^\dagger
    ]
    \nonumber \\
    &
    =   -i\hat{H}_{\rm cond} \hat{\rho}(t) +i\hat{\rho}(t) \hat{H}_{\rm cond}^\dagger
    +
    \sum_\alpha
    [\hat L_\alpha \hat\rho 
    \hat L_\alpha^\dagger
    +
    \hat G_\alpha \hat\rho 
    \hat G_\alpha^\dagger
    ] -\sum_{\alpha,i} |g_i^\alpha|^2 \hat{\rho}(t),
    \label{Eq_Lindlad}
\end{align}
where 
\begin{eqnarray}
\label{H_cond}
    \hat{H}_{\rm cond}:= \hat{H}- \frac{i}{2}  \sum_\alpha ( \hat{L}_\alpha^\dagger \hat{L}_\alpha -\hat{G}_\alpha \hat{G}_\alpha^\dagger)= \hat{H}- \frac{i}{2} ( \hat{L}- \hat{G}).  
\end{eqnarray}
The third term  in Eq.~(\ref{Eq_Lindlad}) is called the jump term, describing the jump process where the number of particles in the relevant system (${\rm S + \bar{S}}$) changes.
In the case where one post-selects the quantum trajectories with no jump processes, the dynamics is governed by a non-Hermitian Hamiltonian $\hat H_{\rm cond}$.

For the unconditional dynamics that we are interested in, the equation of motion of the $(j,i)$ component of the covariance matrix $[{\sf C}]_{j,i}(t)= {\rm Tr}_{\rm S+\bar{S}}[\hat{c}^\dagger_i \hat{c}_j \hat{\rho}(t)]=\langle \hat{c}^\dagger_i \hat{c}_j \rangle$ is obtained by substituting the master equation for the relevant system ${\rm (S+\bar{S})}$ Eq.~(1) in $ i\partial_t \langle \hat{c}^\dagger_i \hat{c}_j \rangle$ such that,
\begin{align}
    i\partial_t [{\sf C}]_{j,i}(t)&= i{\rm Tr}_{\rm S+\bar{S}}[\hat{c}^\dagger_i \hat{c}_j \partial_t\hat{\rho}(t)]\nonumber \\
    &= {\rm Tr}\left[ \hat{c}^\dagger_i \hat{c}_j  \sum_{a,b} H_{a,b} [\hat{c}^\dagger_a \hat{c}_b, \hat{\rho}(t)] +i L_{a,b} \left( \hat{c}_b \hat{\rho}(t) \hat{c}_a^\dagger -\frac{1}{2}\{ \hat{c}^\dagger_a \hat{c}_b, \hat{\rho}(t) \}\right) +i G_{a,b} \left( \hat{c}_a^\dagger \hat{\rho}(t) \hat{c}_b -\frac{1}{2}\{ \hat{c}_b \hat{c}_a^\dagger, \hat{\rho}(t) \}\right) \right] \nonumber \\
    &= \sum_{a,b} H_{a,b} \langle [\hat{c}^\dagger_i \hat{c}_j, \hat{c}_a^\dagger \hat{c}_b] \rangle
    +  \frac{i}{2}\sum_{a,b} \left( L_{a,b}\langle  \hat{c}_a^\dagger [\hat{c}^\dagger_i \hat{c}_j ,\hat{c}_b] +[\hat{c}^\dagger_a, \hat{c}^\dagger_i \hat{c}_j] \hat{c}_b   \rangle +   G_{a,b} \langle  \hat{c}_b [\hat{c}^\dagger_i \hat{c}_j ,\hat{c}_a^\dagger] +[\hat{c}_b, \hat{c}^\dagger_i \hat{c}_j] \hat{c}_a^\dagger   \rangle \right).
\end{align}
Using the algebra of fermionic operators $\{ \hat{c}_i^\dagger, \hat{c}_j \}= \delta_{i,j},\{ \hat{c}_i, \hat{c}_j \}=0  $, we obtain

\begin{align}
     i\partial_t [{\sf C}]_{j,i}(t)&= \sum_a \langle \hat{c}^\dagger_i \hat{c}_a\rangle \left[ H_{j,a} -\frac{i}{2} ( L_{j,a} + G_{j,a})\right] -\sum_a  \langle \hat{c}^\dagger_a \hat{c}_j\rangle \left[ H_{a,i} +\frac{i}{2} ( L_{a,i} + G_{a,i})\right]  +i G_{j,i}.
\end{align}
which is equivalent to Eq.~(5),
where the effective Hamiltonian is given by,
\begin{eqnarray}
    \hat{ H}_{\rm eff}:= {\hat H} -\frac{i}{2}({\hat L}+{\hat G}).
\end{eqnarray}
Therefore, the dynamics of the two point correlation function is governed by $\hat H_{\rm eff}$.
Note the sign difference in front of $\hat G$ compared to the conditional Hamiltonian $\hat H_{\rm cond}$ in Eq.~\eqref{H_cond}.

\subsection{ES and Covariance Matrix Spectrum}
We wish to obtain the relation between the spectrum of the reduced density matrix of the subsystem of 
interest ${\rm S}$ and the spectrum of the covariance matrix. 
As pointed out in the text, for free fermionic systems the reduced density matrix of the subsystem of interest is given by $\hat{\rho}_{\rm S} = {\rm Tr}_{\rm \bar{S}}\hat{\rho}= e^{-\hat{H}_{\rm S}}/{\rm Tr}[e^{-\hat{H}_{\rm S}}]$. Here, $\hat{H}_{\rm S}$ is the 
entanglement Hamiltonian and its spectrum the single particle ES. 
Since the total system Hamiltonian is Gaussian, the reduced density matrix of the subsystem of interest ${\rm S}$
should remain Gaussian. This allows us to generically express the entanglement Hamiltonian as
\begin{equation}
    \hat{H}_{\rm S} = \sum_{n,m} h\mathsf{}_{n,m}\hat{c}^\dagger_n \hat{c}_m.
    \label{H_ent}
\end{equation}
Using Eq.~(\ref{H_ent}), the covariance matrix elements of the subsystem S are given by  
\begin{align}
     [{\sf C_{\rm SS}}]_{m,n}
     &={\rm Tr}_{\rm S}[\hat{c}^\dagger_n \hat{c}_m \hat{\rho}_{\rm S}]= \frac{1}{Z_{\rm S}} {\rm Tr_{\rm S}}[\hat{c}^\dagger_n \hat{c}_m e^{-\hat{H}_{\rm S} }] \;\;(n,m \in {\rm S}).
     \label{SScov}
\end{align}
Let $\{ |\phi_l\rangle = f^\dagger_l|0\rangle \}$
be a basis that diagonalizes the entanglement Hamiltonian, such that $ \hat{H}_S= \sum_l \epsilon_l \hat{f}^\dagger_l \hat{f}_l $,
where
$\hat{c}_n = \sum_l  \phi_l(n) \hat{f}_l$ 
and $\hat{c}_n| 0\rangle= \hat{f}_l| 0\rangle=0$. Substituting on Eq.~(\ref{SScov}),  we obtain
\begin{align}
    [{\sf C}_{\rm SS}]_{m,n}&=
    {\rm Tr} \left[\sum_{\mu,\nu} \phi_\mu^*(n) \phi_\nu(m)\hat{f}^\dagger_\mu \hat{f}_\nu \exp( -\sum_l \epsilon_l \hat{f}_l^\dagger \hat{f}_l )\right]\Bigg/  {\rm Tr} \left[\exp( -\sum_l \epsilon_l \hat{f}_l^\dagger \hat{f}_l )\right], \\
   &= \sum_l \phi_l(n)^* \phi_l(m) \frac{1}{e^{\epsilon_l}+1}.
    \label{SubsystemCov}
\end{align}
Diagonalizing ${ \sf C}_{\rm SS}$ and using Eq.~(\ref{SubsystemCov}) we obtain Eq.~(4).
\subsection{ES crossings are determined by the topology of $\hat H_{\rm eff}$ under OBC and their robustness against disorder and local dissipation }

\begin{figure}[h!]
    \centering
    \includegraphics[width=0.9\textwidth]{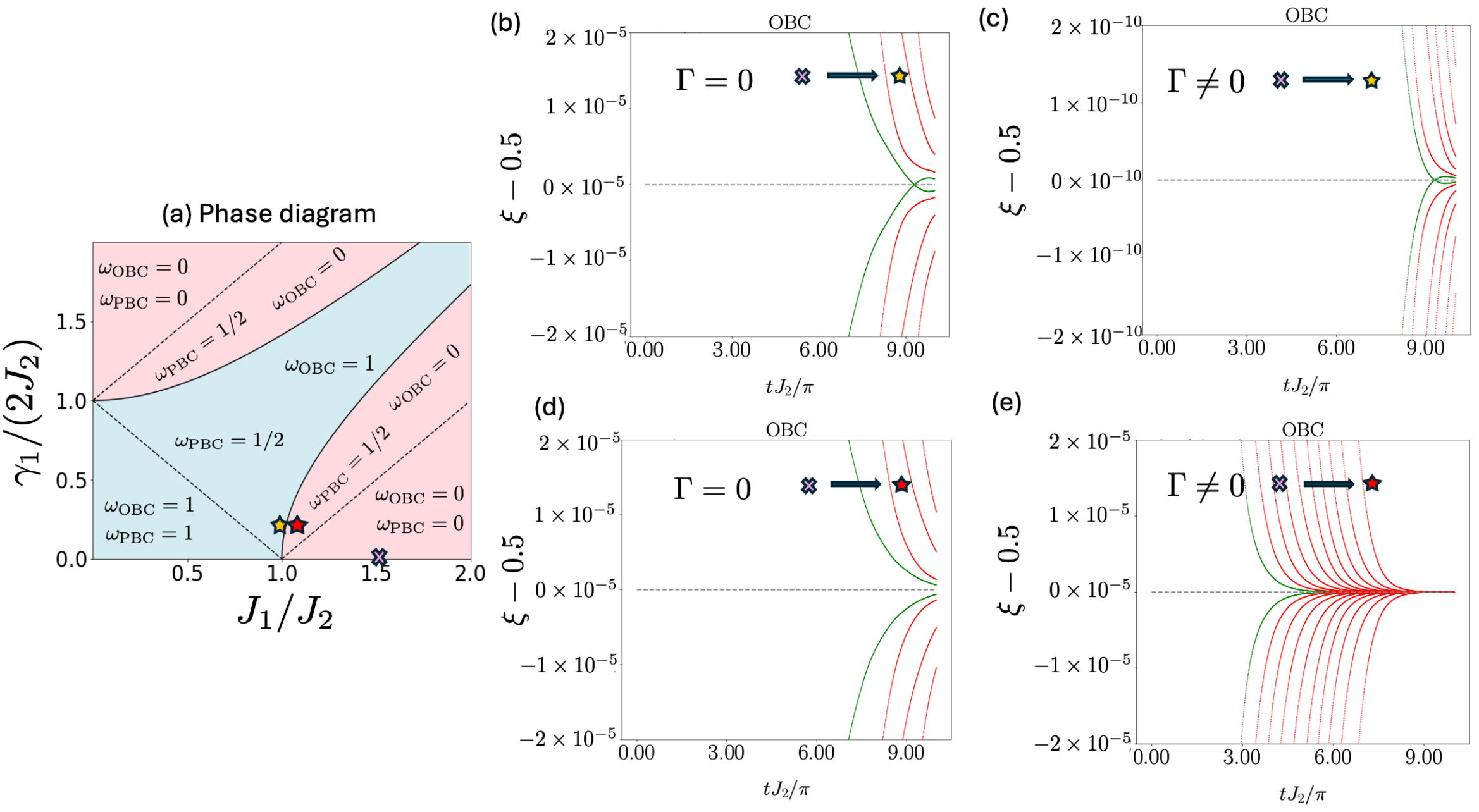}
    \caption{
      Quench dynamics of the ES under OBC. The system is initially prepared in the trivial ground state of the SSH model ~[S37] $(J_1/J_2, \gamma_1/(2J_2))=(1.5,0)$. Here we consider a system of 60 sites and a 10 sites subsystem S. (a) Phase diagram of the NH SSH model Eq.~(7).
Topological phase transitions are separated by the solid and dotted lines for OBC and PBC spectra, respectively. (b)-(e), Quench dynamics of ES under OBC. The green lines represent the two eigenvalues, $\xi_l(t)$, of the covariance matrix $\mathsf{C}_{\rm SS}(t)$ that are closest to the zero energy entanglement value $\xi=0.5$ during the time evolution.  Other eigenvalues are shown in red. (b) Trivial to topological phase quench without uniform dissipation
    $(J_1/J_2, \gamma_1/(2J_2), \Gamma/(2J_2))= (1.009,  0.2, 0)$. A pair of ES eigenvalues cross at $\xi=0.5$ while decaying. 
(c) Trivial to topological phase  quench with uniform dissipation
    $(J_1/J_2, \gamma_1/(2 J_2), \Gamma/(2 J_2))= (1.009,  0.2, 0.1)$. A pair of ES eigenvalues cross at $\xi=0.5$ while decaying. (d) Trivial to trivial phase quench without uniform dissipation $(J_1/J_2, \gamma_1/(2J_2), \Gamma/(2J_2))= (1.029,  0.2, 0)$. All eigenmodes decay to $\xi=0.5$ without crossing it. (e) Trivial to trivial phase quench with uniform dissipation $(J_1/J_2, \gamma_1/(2J_2), \Gamma/(2J_2) )= (1.029,  0.2, 0.1)$. All eigenmodes decay to $\xi=0.5$ without crossing it. } 
    \label{fig_boundary_OBC}
\end{figure}

\begin{figure}[h]
    \centering
    \includegraphics[width=0.9\textwidth]{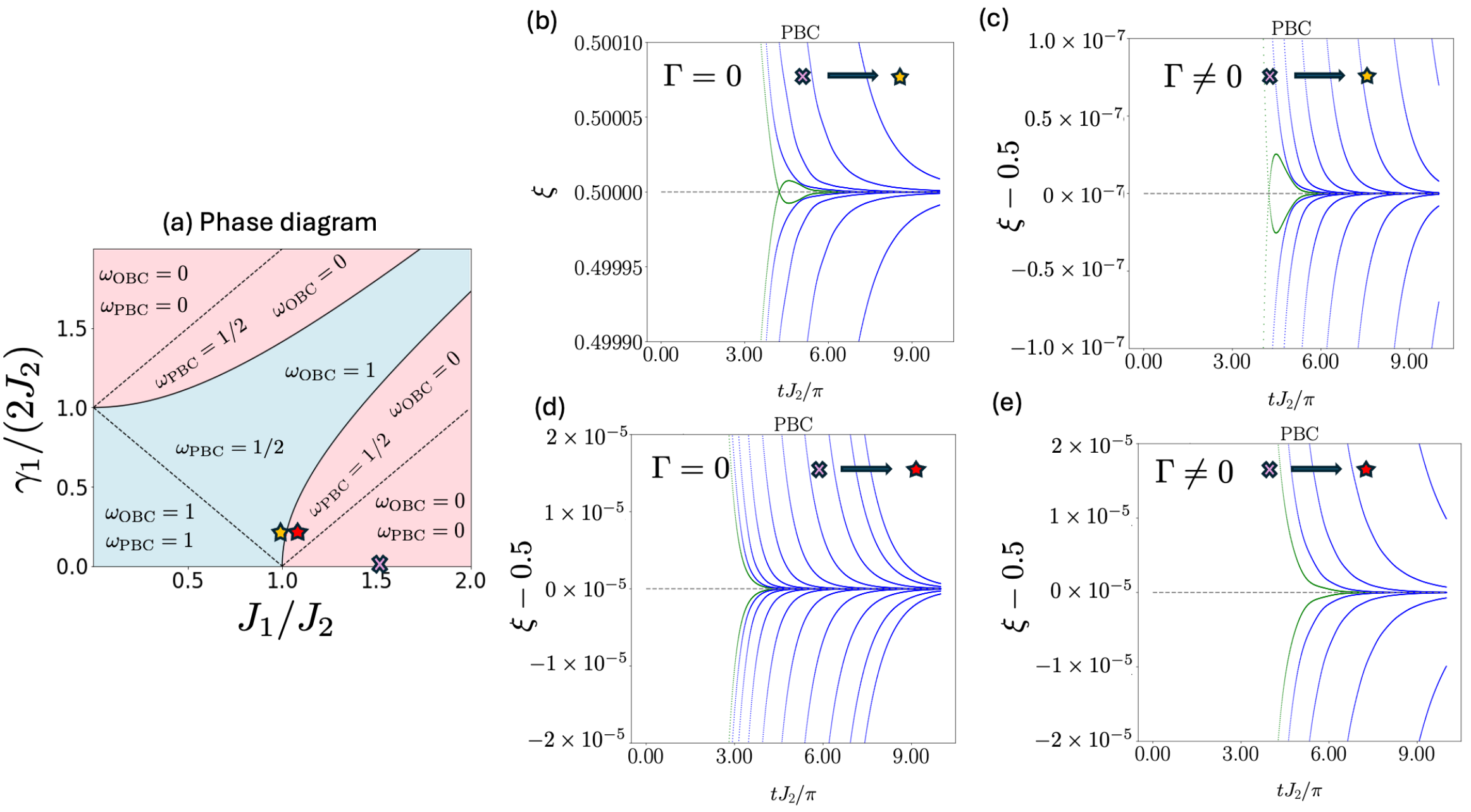}
    \caption{  Quench dynamics of the ES under PBC. The system is initially prepared in the trivial ground state of the SSH model ~[S37] $(J_1/J_2, \gamma_1/(2J_2))=(1.5,0)$. Here we consider a system of 60 sites and a 10 sites subsystem S. (a) Phase diagram of the NH SSH model Eq.~(7).
Topological phase transitions are separated by the solid and dotted lines for OBC and PBC spectra, respectively. (b)-(e), Quench dynamics of ES under PBC.  The green lines represent the two eigenvalues, $\xi_l(t)$, of the covariance matrix $\mathsf{C}_{\rm SS}(t)$ that are closest to the zero energy entanglement value $\xi=0.5$ during the time evolution.  Other eigenvalues are shown in blue. (b) Trivial to topological phase quench without uniform dissipation
    $(J_1/J_2, \gamma_1/(2J_2), \Gamma/(2J_2))= (1.009,  0.2, 0)$. A pair of ES eigenvalues cross at $\xi=0.5$ while decaying.
(c) Trivial to topological phase  quench with uniform dissipation
    $(J_1/J_2, \gamma_1/(2 J_2), \Gamma/(2 J_2))= (1.009,  0.2, 0.1)$. A pair of ES eigenvalues cross at $\xi=0.5$ while decaying. (d) Trivial to trivial phase quench without uniform dissipation $(J_1/J_2, \gamma_1/(2J_2), \Gamma/(2J_2) )= (1.029,  0.2, 0)$. All eigenmodes decay to $\xi=0.5$ without crossing it. (e) Trivial to trivial phase quench with uniform dissipation $(J_1/J_2, \gamma_1/(2J_2), \Gamma/(2J_2))= (1.029,  0.2, 0.1)$. All eigenmodes decay to $\xi=0.5$ without crossing it.} 
    \label{fig_boundary_PBC}
\end{figure}

Here, we numerically confirm that the ES crossings are determined by the topology of $\hat H_{\rm eff}$ under OBC and are robust against disorder and local dissipation. 
Figures \ref{fig_boundary_OBC} and \ref{fig_boundary_PBC} and show the ES dynamics after a dissipative quench under OBC and PBC, respectively. Post-quench parameters are chosen to be very close to the phase boundary of the topological transition of the OBC spectrum shown in panel (a) of Figures \ref{fig_boundary_OBC} and \ref{fig_boundary_PBC}.
Here, as in the situation considered in the main text, the pre-quench state is chosen to be in a topologically trivial state. 
As shown, zero-crossings of the ES (does not) occur when the post-quench parameters are in the topological (trivial) phase of $\hat{H}_{\rm eff}$, regardless of the boundary conditions of the system ${\rm S+\bar{S} }$.
This demonstrates that the ES crossing probes the non-Hermitian bulk-boundary correspondence in a system under periodic or open boundary conditions.
Table~\ref{Table_Boundaries} summarizes the presence (or absence) of ES crossings based on the topological phase of the effective Hamiltonian, $\hat{H}_{\rm eff}$, when starting from a trivial ground state under both open and periodic boundary conditions.\\
\begin{table}[h!]
    \centering
\begin{tabular}{|c|c|c|c|c|}
\hline
System Boundary Condition & Initial State Phase & Effective Hamiltonian Phase & ES Crossings & Figure \\ 
\hline
OBC & Trivial & Topological & Present & S1(b) \\ 
\hline
OBC & Trivial & Trivial & Absent & S1(d) \\ 
\hline
PBC & Trivial & Topological & Present & S2(b) \\ 
\hline
PBC & Trivial & Trivial & Absent & S2(d) \\ 
\hline

\end{tabular}
\caption{Presence or absence of entanglement spectrum crossings for different boundary conditions, initial phases, and quench phases.}
\label{Table_Boundaries}
\end{table}
\\

Furthermore, we confirm below that this property is robust against local dissipation.
This is motivated by the property that was pointed out recently~\cite{Lee23, McDonald22, Mao21} that the presence of the LSE generically does 
\textit{not} imply that correlation functions have sensitivity to boundary conditions,
especially in the presence of a local dissipation~\cite{Lee23, McDonald22}. 
Intuitively, this can be understood as follows: a wave packet prepared in the middle of the bulk would not know about the boundary condition unless it hits the boundary. However, as long as the wave packet has a finite lifetime, it generically decays before it propagates to the edge. 

\begin{figure}[h!]
    \centering
    \includegraphics[width=0.6\textwidth]{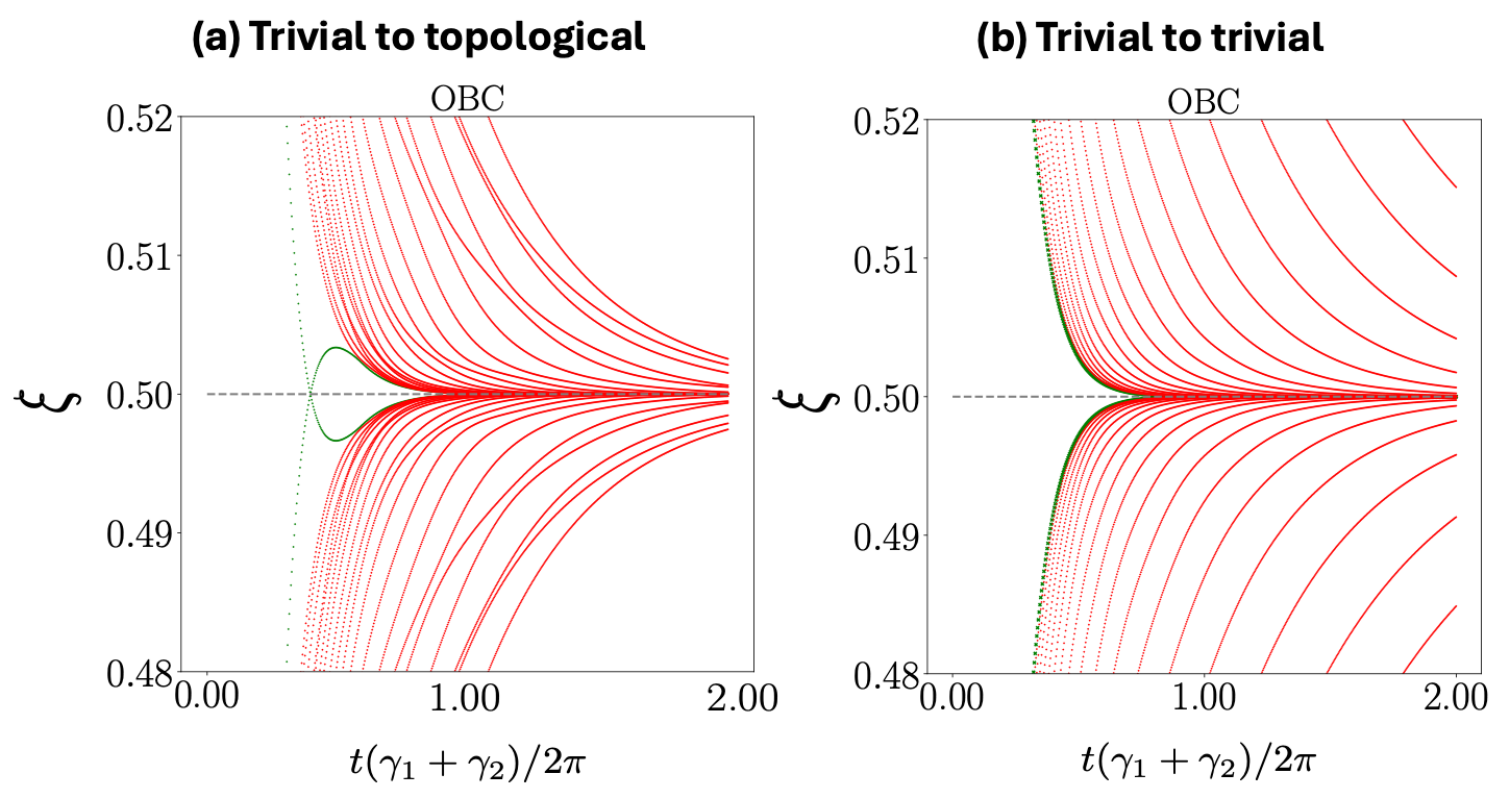}
    \caption{
    (a),(b) Quench dynamics of ES under OBC. The system is initially prepared in the trivial ground state of the SSH model $(J_2/J_1, J_3/J_1, \gamma_1/J_1,\gamma_2/J_1)= (0,0,0,0)$, 
    with system size $L$ = 40, and a subsystem S
    of $M=20$ sites. At $t=0$ the relevant system $(\mathrm{S}+\bar{\mathrm{S})}$ is driven out of equilibrium by an abrupt change in parameters and allowing the system to couple to Markovian reservoirs at a rate $\gamma_i, (i=1,2)$. The green lines represent the two eigenvalues, $\xi_l(t)$, of the covariance matrix $\mathsf{C}_{\rm SS}(t)$ that are closest to $\xi=0.5$ during the time evolution.  The other eigenvalues are shown in red. (a) Trivial to topological quench for OBC and PBC spectrum $(2J_1/(\gamma_1+\gamma_2),2J_2/(\gamma_1+\gamma_2),2J_3/(\gamma_1+\gamma_2))=(-1,1.4,1/5)$.  A pair of ES eigenvalues, $\xi_l(t)$, cross $\xi=0.5$ as they decay. (b) Trivial to trivial (topological) quench for OBC (PBC) spectrum $(2J_1/ (\gamma_1+\gamma_2),2J_2/(\gamma_1+\gamma_2),2J_3/(\gamma_1+\gamma_2))=(2.1,1.4,1/5)$.  All the eigenvalues $\xi_l(t)$ decay without zero-crossings. }
    \label{PBC_OBC_J3}
\end{figure}
\pagebreak
A natural question to ask is whether ES zero-crossings are also sensitive to such perturbations, which we confirmed negative.
\begin{figure}[h!]
    \centering
    \includegraphics[width=0.6\textwidth]{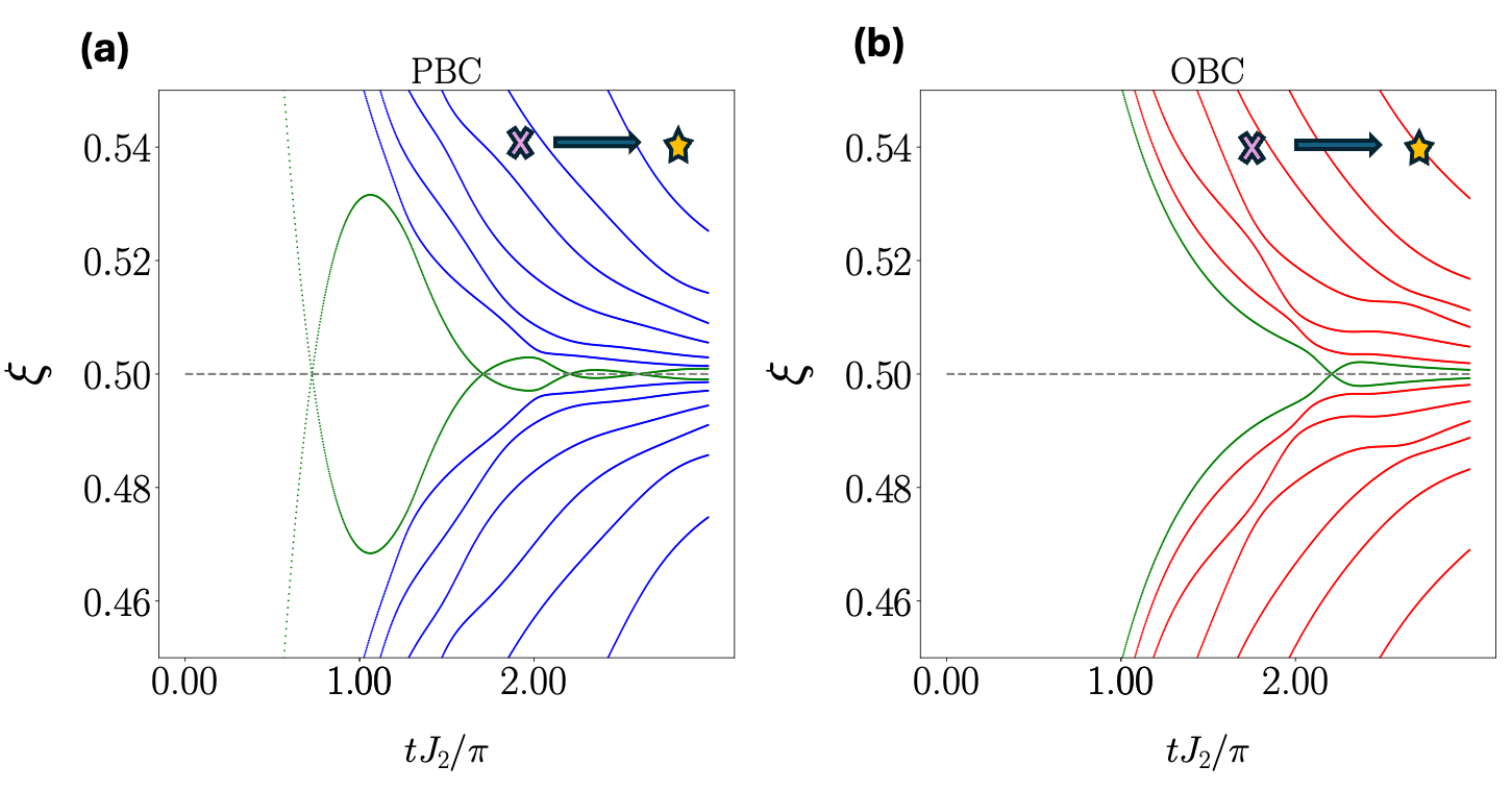}
    \caption{
    The system is initially prepared in the trivial ground state of the SSH model under PBC for (a), and under OBC for (b), with no dissipation.
        Then, at $t >0$, the relevant system $(\mathrm{S}+\bar{\mathrm{S}})$ is driven out of its initial state by an abrupt change of the random hopping parameters $J_{1,j}, J_{2,j}$ and a correlated dissipation with rate $\gamma_1$ is simultaneously turned on. The relevant system evolves according to the GKLS equation 
        under PBC for (a), and under OBC for (b), respectively.
        (a), (b) ES dynamics for a subsystem S with random perturbations that preserve the symmetries of the effective Hamiltonian. (a) Quench dynamics of ES for
        $(\bar{J}_1/\bar{J}_2,\gamma_1/(2\bar{J}_2))=(0.7,0.2)$. The system evolves under PBC. The two eigenvalues of the covariance matrix $\mathsf{C}_{\rm SS}(t)$ closest to $\xi_l(t)=0.5$ are shown in green. Other eigenvalues are shown in blue. One pair of ES eigenvalues
        cross $\xi$ = 0.5 as they decay. 
        (b) Quench dynamics of ES for 
         $(\bar{J}_1/\bar{J}_2,\gamma_1/(2\bar{J}_2))=(0.7,0.2)$. The system evolves under OBC. 
        The two  eigenvalues of the covariance matrix $\mathsf{C}_{\rm SS}(t)$ closest to $\xi_l(t)=0.5$ are shown in green. Other eigenvalues are shown in red. One pair of ES eigenvalues
        cross $\xi$ = 0.5 as they decay. 
        We set the initial state to be the (trivial) ground state at $(J_1/J_2, \gamma_1/(2J_2))=(1.5,0)$,
        the system size $L=20$, and the size of subsystem S is $M=10$ sites. In the clean limit, $J_{1,j}=\bar{J}_1, J_{2,j}=\bar{J}_2$, the quench Liouvillian lies on a topological phase for both OBC and PBC spectrum ($\omega_{\rm OBC}=1, \omega_{\rm PBC}=1$). Note crucially that the ES dynamics reflect the OBC spectrum, regardless whether the physical system is in PBC and whether the system is subject to random perturbations which preserve the symmetries of the effective Hamiltonian.}
    \label{Dirty_Clean}
\end{figure}
Panels (c) and (e) of Figures~\ref{fig_boundary_OBC} and~\ref{fig_boundary_PBC} show the effects of adding local gain and loss with rate $\Gamma$ to the ES dynamics under OBC and PBC, respectively. The jump operators are given by $\hat{L}_{A(B), j}^{'}:=\sqrt{ \Gamma/2} \;\hat{c}_{A(B), j}$ and $\hat{G}_{A(B), j}^{'}:=\sqrt{ \Gamma/2}\; \hat{c}_{A(B),j}^\dagger$ . 
As these onsite gains and losses are local, the added terms merely shift the effective Hamiltonian $\hat H_{\rm eff}$ in the imaginary axis and therefore do not change its topological properties.
Here, we have assumed that the added loss and gain are balanced, to ensure that the system converges to an infinite temperature state, for the same reason we describe in the main text (See also~\cite{Sayyad21,NOTE}). 
We see that even when we perturb the system with these series of local system-environment interactions, the property that the presence of ES zero-crossings 
reflects the topology of $\hat H_{\rm eff}$ remain robust even near the phase boundary. Interestingly, the presence or absence of ES crossings occurs in both PBC and OBC dynamics. However, the primary difference is that, under OBC dynamics, the time required for the first ES crossing is significantly larger compared to PBC dynamics.
A similar qualitative behavior is observed in quenches where the spectrum of the effective Hamiltonian cannot be obtained through a similarity transformation ($J_3,\gamma_2 \neq 0$), as depicted in Figure~\ref{PBC_OBC_J3}.   

To demonstrate that ES crossings have a topological origin, we introduced disorder to the effective Hamiltonian in two ways. 
First, we add disorder that preserves the
sublattice symmetry as follows:

\begin{align}
    \hat{H}_{\rm eff}^{\rm (A)} &= \sum \limits_{j} \Big[ \Big(J_{1,j} \hat{c}_{A,j}^\dagger \hat{c}_{B,j} + J_{2,j} \hat{c}^\dagger_{B,j} \hat{c}_{A,j+1} + J_{3,j} \hat{c}_{A,j}^\dagger \hat{c}_{B,j+1} \Big) + h.c. \Big]  \nonumber \\
    &+ \sum_j \Big( \frac{\gamma_1}{2}( \hat{c}_{A,j}^\dagger \hat{c}_{B,j} - \hat{c}^\dagger_{B,j} \hat{c}_{A,j} ) + \frac{\gamma_2}{2}( \hat{c}^\dagger_{B,j} \hat{c}_{A,j+1} - \hat{c}_{A,j+1}^\dagger \hat{c}_{B,j}) \Big) - i \frac{\gamma_1 + \gamma_2}{2} \hat{n},
\end{align}
where $J_{\alpha,j}$ are randomly sampled from a uniform distribution over $[0.6 \bar{J}_{\alpha}, 1.4 \bar{J}_{\alpha}]$.
This preserves the sublattice symmetry $\{\hat H_{\rm eff}^{(A)},\hat S\}=0$, where $\hat S$ satisfies $\hat{S}\hat{\bf c}(k,k') \hat{S}^{-1}= \sigma_z\hat{\bf c}(-k, -k') $, $\boldsymbol{c}(k) := (\hat{c}_A(k), \hat{c}_B(k'))^T$, and we ignored the constant shift term.

\begin{figure}[h!]
    \centering
    \includegraphics[width=0.6\textwidth]{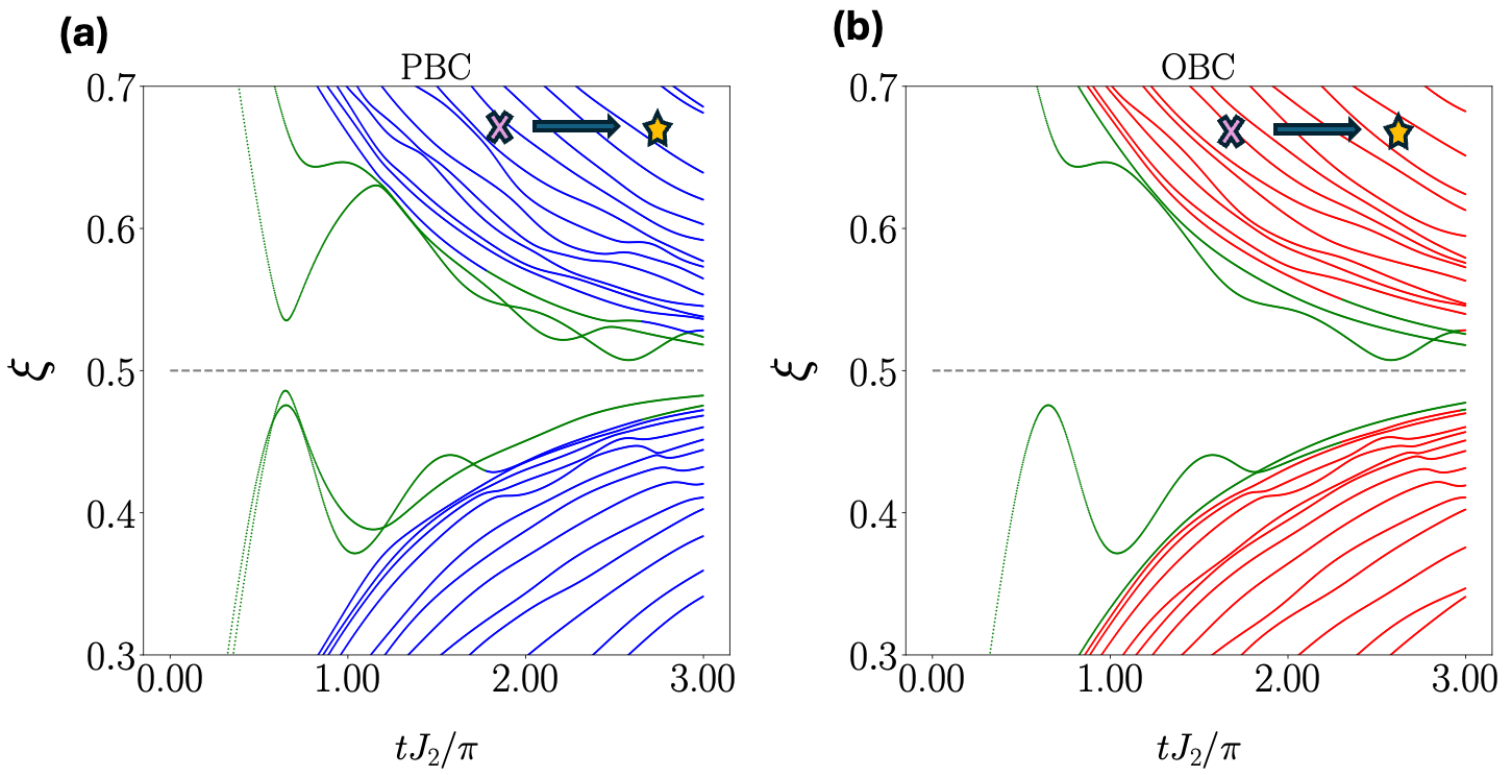}
    \caption{ The system is initially prepared in the trivial ground state of the SSH model, under PBC for (a), and under OBC for (b), with no dissipation.
        Then, at $t >0$, the relevant system $(\mathrm{S}+\bar{\mathrm{S}})$ is driven out of its initial state by an abrupt change of the hopping parameters $J_{1}, J_{2}$, with a random on-site potential $W_j \in [-W,W]$, and a correlated dissipation with rate $\gamma_1$ is simultaneously turned on. The relevant system evolves according to the GKLS equation 
        under PBC for (a), and under OBC for (b), respectively.
        (a),(b) ES dynamics for a subsystem S with random onsite disorder. We set the disorder strength to $W=0.8$. (a) Quench dynamics of ES for $(J_1/J_2,\gamma_1/(2J_2))=(0.5,0.1)$.  The system evolves under PBC. 
        The four eigenvalues of the covariance matrix $\mathsf{C}_{\rm SS}(t)$ closest to $\xi_l(t)=0.5$ are shown in green. Other eigenvalues are shown in blue. No ES eigenvalues
        cross $\xi$ = 0.5 as they decay. 
        (b) Quench dynamics of ES for $(J_1/J_2,\gamma_1/(2J_2))=(0.5,0.1)$. The system evolves under OBC.
        The four eigenvalues of the covariance matrix $\mathsf{C}_{\rm SS}(t)$ closest to $\xi_l(t)=0.5$ are shown in green. Other eigenvalues are shown in red. No ES eigenvalues cross $\xi$ = 0.5 as they decay.  We set the initial state to be the (trivial) ground state at $(J_1/J_2, \gamma_1/(2J_2))=(1.5,0)$,
        the system size $L=68$, and the size of subsystem S is $M=34$ sites. In the clean limit, $W_j=0$, the quench Liouvillian lies on a topological phase for both OBC and PBC spectrum ($\omega_{\rm OBC}=1, \omega_{\rm PBC}=1$). Note that no ES crossings are present when the system is subject to random perturbations which do not preserve the symmetries of the effective Hamiltonian.}
    \label{Dirty_CS}
\end{figure}

Second, we introduce disorder that explicitly breaks the sublattice symmetry using a random on-site potential. The modified effective Hamiltonian in this case is given by:
\begin{align}
    \hat{H}_{\rm eff}^{\rm (B)} 
    &= \sum \limits_{j} \Big[ \Big(J_{1} \hat{c}_{A,j}^\dagger \hat{c}_{B,j} + J_{2} \hat{c}^\dagger_{B,j} \hat{c}_{A,j+1} + J_{3} \hat{c}_{A,j}^\dagger \hat{c}_{B,j+1} \Big) + h.c. \Big]  \nonumber \\
    &+ \sum_j \Big( \frac{\gamma_1}{2}( \hat{c}_{A,j}^\dagger \hat{c}_{B,j} - \hat{c}^\dagger_{B,j} \hat{c}_{A,j} ) + \frac{\gamma_2}{2}( \hat{c}^\dagger_{B,j} \hat{c}_{A,j+1} - \hat{c}_{A,j+1}^\dagger \hat{c}_{B,j}) \Big) - i \frac{\gamma_1 + \gamma_2}{2} \hat{n} \nonumber \\
    &+ \sum_j W_j (\hat{c}_{B,j}^\dagger \hat{c}_{B,j} + \hat{c}_{A,j}^\dagger \hat{c}_{A,j}),
\end{align}
where $W_j$ are randomly sampled from a uniform distribution over $[-W, W]$. Sublattice symmetry is broken since $\{\hat{S},\sum_j W_j (\hat{c}_{B,j}^\dagger \hat{c}_{B,j} + \hat{c}_{A,j}^\dagger \hat{c}_{A,j})\}\neq 0$.
As shown in Figure~\ref{Dirty_Clean}, our findings reveal that entanglement spectrum (ES) crossings remain robust during a quench from a trivial phase to a topological phase of the effective Hamiltonian when the disorder respects the symmetries of the effective Hamiltonian ($\hat H_{\rm eff}^{\rm (A)}$). In contrast, as illustrated in Figure~\ref{Dirty_CS} , ES crossings disappear when the disorder breaks sublattice symmetry ($\hat H_{\rm eff}^{\rm (B)}$).

\section*{Effective  Hamiltonian for the Lindbladian SSH model}
Here we give explicit expressions for the non-Hermitian effective Hamiltonian Eq.~(2) which determines the dynamics of the covariance matrix. As mentioned earlier, we focus on the case where the matrices $\mathsf{L},\mathsf{G}$ corresponding to $\hat{L}$ and $\hat{G}$ satisfy $\mathsf{L} = \mathsf{G}$. Hence, the steady state is given by the infinite temperature state $C(\infty) = (1/2) \mathsf{I}_{2N,2N}$, where $\mathsf{I}_{2N,2N}$ is the unit matrix of size $2N\times 2N$. From this, we can readily see that all $\xi_l $ converge to $0.5$ at $t \rightarrow \infty$. The formal solution of Eq.~(5) is given by 
\begin{equation}
     \mathsf{C}(t)= \mathsf{C}(\infty) + e^{-i \mathsf{H}_{\rm eff} t}
    \Big( \mathsf{C}(0)- \mathsf{C}(\infty)\Big) e^{i \mathsf{H}_{\rm eff}^\dagger t}.
    \label{covariance_fast}
\end{equation}
Due to the Gaussian nature of the initial state and its time evolution, the ES can be straightforwardly obtained by diagonalizing the covariance matrix restricted to the region S, $\mathsf{C}_{\rm SS}$. 
The effective (non-Hermitian) Hamiltonian reads
\begin{equation}
    {\sf H}_{\rm eff}= \begin{pmatrix}
        -i(\gamma_1+\gamma_2)/2    & J_1+\gamma_1/2 & 0                   & J_3                   & 0      & 0      & \dots   \\[5pt]
        J_1- \gamma_1/2 & -i(\gamma_1+\gamma_2)/2  & J_2+\gamma_2/2                 & 0                   & 0      & 0      & \dots   \\[5pt]
        0                    & J_2-\gamma_2/2                 & -i(\gamma_1+\gamma_2)/2   & J_1+ \gamma_1/2 & 0      & J_3      & \ddots  \\[5pt]
        J_3                    & 0                   & J_1-\gamma_1/2 & -i(\gamma_1+\gamma_2)/2   & J_2+\gamma/2    & 0      & \ddots  \\
        \vdots               & \vdots              & \ddots              & \ddots              & \ddots & \ddots & \ddots
    \end{pmatrix}.
    \label{HeffMatrix}
\end{equation}
To obtain the ES of the subsystem S we numerically evaluate ${\sf C}(t)$ 
and obtain $\xi_l$ from the spectrum of ${\sf  C}_{\rm SS}(t)$. 
The initial covariance matrix  ${\sf C}(0)$ is given by $[{\sf C}]_{j,i}(0)= \langle GS |\hat{c}^\dagger_i \hat{c}_j |GS \rangle$, where $|GS\rangle$ is the ground state of the relevant system $\sf (S+\bar{S})$ in the Hermitian limit. 
For translationally invariant systems, the many-body ground state wavefunction $|GS\rangle$ is a single Slater determinant of all single-particle eigenstates of $\hat{H}$ with energies less than the Fermi energy $\epsilon_F=0$, $|GS \rangle= \Pi_{k;\epsilon(k) <0} \sum_{\alpha} (u_k^\alpha c^\dagger_{\alpha}(k))|0\rangle$, where $\ket{0}$ is the vacuum state, $|\alpha,k\rangle= c^\dagger_\alpha(k)|0 \rangle$,  $u^\alpha_k= \langle \alpha,k|\psi(k) \rangle (\alpha\in\{A,B\})$, and  $\hat{H}|\psi(k) \rangle= \epsilon(k)|\psi(k) \rangle$.


\begin{thebibliography}{99}
\bibitem{NOTE}
    For example,
    consider a system which has loss only (i.e., $\hat G_{j,s=1,2}=0$)
    with the initial state given by an all-occupied state.
    In this case, the system keeps losing particles from the system and would eventually converge to a vacuum state. 
    This situation gives rise to the ES crossing for a trivial reason:
    the ES at the initial state $\xi_l(t=0)=1$ converge to $\xi_l(t\rightarrow\infty)=0$, which gives rise to ES crossing $\xi_l(t)=0.5$ at finite time $t$, irrespective of the topology of $\hat H_{\rm eff}$.
\end{thebibliography}
\end{document}